\documentclass[doublecol]{epl2}
\usepackage{pifont}
\usepackage{txfonts}
\usepackage{amssymb}
\usepackage{graphicx}
\usepackage{amsfonts}
\usepackage{epsfig}

 \institute{
  \inst{1} Department of Physics and ITP, The Chinese University of Hong Kong,
  Hong Kong, China\\
  \inst{2} Institute of Theoretical Physics, Chinese Academy of Sciences,
  Beijing, 100080, China}
\abstract{Using the numerical unrestricted Hartree-Fock approach, we study the
ground state of a two-orbital model describing newly discovered FeAs-based
superconductors. We observe the competition of a $(0, \pi)$ mode spin-density
wave and the superconductivity as the doping concentration changes. There might
be a small region in the electron-doping side where the magnetism and
superconductivity coexist. The superconducting pairing is found to be spin
singlet, orbital even, and mixed s$_{xy}$ + d$_{x^{2}-y^{2}}$ wave (even
parity).}

\begin{document}

\title{Competitions of magnetism and superconductivity in FeAs-based
materials}
\author{Shuo Yang\inst{1,2} \and Wen-Long You\inst{1} \and Shi-Jian Gu%
\inst{1} \and Hai-Qing Lin\inst{1}}
\maketitle

\shortauthor{S. Yang \etal}

The newly discovered FeAs-based superconductors \cite%
{KamiharaJACS,KakahashiNature} have attracted lots of experimental\cite%
{HHWen,NLWang,ZXZhao,DFFang08033603,DLFeng08034328,PCDai08040795,ZhuanXu}
and theoretical \cite%
{DJSingh08030429,KHaule08031279,GXu08031282,LBoeri08032703,IIMazin08032740,
CCao08033236,FJMa08033286,ZYLu08033286,FJMa08043370,QHChen08050632,
XDai08033982,ZDWang08034346,TLi08040536,
arXiv08041113,PALee08041739,DXYao08044115,arXiv08042252,QMSi08042480,
ZYWeng08043228,ZJYao08044166,XLQi08044332,MDaghofer08050148,
YPWang08050644,DHLee08053343,YRan08053535,JRShi08060259,
JPHuarXiv08052958,ZHWang08050736,YWanarXiv08050923,YZhou08060712} interests.
Experimentally, the superconducting transition temperature, up to now, can
be as high as 56K\cite{ZXZhao,NLWang,ZhuanXu}. Spin-density wave order of $%
(0, \pi)$ mode was observed in the parent compound LaOFeAs, but vanishes at
high temperature (above 150K) and large doping region\cite%
{NLWang,PCDai08040795}. However, unlike the cuprate high-Tc superconductors
whose parent compound is an insulator, LaOFeAs is a semimetal\cite%
{NLWang,DJSingh08030429}. The observed magnetic dependence of the specific
heat as well as the results of nuclear magnetic resonate suggest the
presence of gapless nodal lines on the Fermi surface\cite{HHWen}.
Theoretically, the transition temperature estimated based on the
electron-phonon coupling seems unlikely to explain the observed
superconductivity, thus suggesting these materials might be unconventional
and non-electron-phonon mediated \cite{LBoeri08032703}. The
local-density-approximation (LDA) calculations show that the density of
state near the Fermi surfaces of the parent compound LaOFeAs are dominated
by iron's 3d electrons\cite%
{DJSingh08030429,KHaule08031279,GXu08031282,IIMazin08032740,CCao08033236,
FJMa08033286}. These observations imply that the multi-orbital effects play
a key role in these new family of high-Tc superconductors. Though irons in
LaOFeAs have five orbitals and ten bands, some groups proposed that it is
sufficient to consider only a few of them, say $d_{xz}$ and $d_{yz}$
orbitals for a minimal two-band model or may include $d_{xy}$ orbital for a
three-band model, to reproduce qualitatively the LDA Fermi surface topology%
\cite{XDai08033982,ZDWang08034346,TLi08040536,arXiv08041113,PALee08041739,
QMSi08042480,ZYWeng08043228,ZJYao08044166,XLQi08044332,MDaghofer08050148}.
The spin-density-wave (SDW) of $(0, \pi)$ mode of the parent compound
LaOFeAs has been interpreted due to the superexchange interaction between
the next-nearest neighboring sites, which leads to an effective $J_1-J_2$
model defined on a two-dimensional lattice \cite%
{DXYao08044115,arXiv08042252,FJMa08043370,QMSi08042480,JPHuarXiv08052958}.
Alternatively, the SDW mode was also attributed to the nesting of Fermi
surface \cite{NLWang,GXu08031282,ZDWang08034346,ZJYao08044166,XLQi08044332,
DHLee08053343,YRan08053535}. On the other hand, despite of the pairing
mechanism being unclear, some groups \cite%
{ZHWang08050736,YWanarXiv08050923,YZhou08060712,JRShi08060259} have
addressed the classification of various superconducting states of the
systems with two orbitals based on group theory.

In this paper, we study the competition between superconductivity and
magnetism, as well as the superconducting pairing symmetry in the FeAs-based
materials. Based on a recently proposed two-orbital model, we introduce the
superconducting pairing into the Hamiltonian from which various pairing
possibility can be constructed. Then we study the ground state property by
using the unrestricted Hartree-Fock approach. We start with random values
for all possible order parameters and let the self-consistent equations
converge to the final solution. We find that for the parent compound, there
exists a $(0, \pi)$ mode spin-density-wave. When doping is introduced,
either by electron or hole, superconductivity appears with the disappearance
of the magnetism, and there is a small overlap in electron-doping region
where the magnetism and superconductivity might coexist. Therefore, these is
a competition between the two orders. In the superconducting region, a
careful scrutiny reveals that the pairing symmetry of the superconductivity
might involve spin singlet, orbital symmetric, and mixed s$_{xy}$+d$_{x^2
-y^2}$ wave (even parity).

\begin{figure}[tbp]
\includegraphics[bb=55 485 535 750, width=7cm, clip] {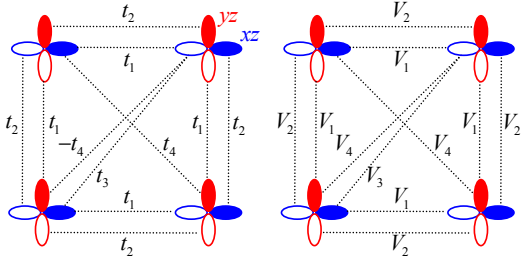}
\caption{ (color online) The schematic illustration for the hopping
parameters and the attractive interaction parameters of the two-orbital $%
d_{xz}-d_{yz}$ model on a square lattice.}
\label{hopping}
\end{figure}

In order to simplify the calculation, we adopt the two-band model of the Fe
subsystem defined on the principle plane, as suggested by Raghu \emph{etal}
\cite{arXiv08041113}, as a starting model for discussing the interplay
between magnetism and superconductivity. While there are arguments as to
whether the minimum model should consist of three orbitals, we expect no
qualitative difference to appear within our calculation framework. There are
two \textquotedblleft $d_{xz}$, $d_{yz}$" orbitals on each site. The left
plot of fig. \ref{hopping} shows all possible hoping processes between
neighboring sites and orbitals based on the tight binding approximations.
Phenomenologically, we use the following mean-field Hamiltonian,%
\begin{eqnarray}
H &=&H_{0}+H_{SC}+H_{I}, \\
H_{0} &=&\sum_{k\sigma }\psi _{k\sigma }^{\dag }\left[ \epsilon _{k}\tau
_{0}+\gamma _{k}\tau _{1}+\xi _{k}\tau _{3}\right] \psi _{k\sigma }, \\
H_{I} &=&-U\sum_{i,\beta }m_{i}^{\beta }\left( C_{i,\uparrow }^{\beta \dag
}C_{i,\uparrow }^{\beta }-C_{i,,\downarrow }^{\beta \dag }C_{i,\downarrow
}^{\beta }\right)   \nonumber \\
&&+\sum_{i,\sigma }\left( U^{\prime }+\frac{J_{H}}{4}\right) \left( \rho
_{i,\sigma }^{a}n_{i,-\sigma }^{b}-\frac{1}{2}\rho _{i,\sigma }^{a}\rho
_{i,-\sigma }^{b}\right)   \nonumber \\
&&+\sum_{i,\sigma }\left( U^{\prime }-\frac{J_{H}}{4}\right) \left( \rho
_{i,\sigma }^{a}n_{i,-\sigma }^{b}-\frac{1}{2}\rho _{i,\sigma }^{a}\rho
_{i,\sigma }^{b}\right)   \nonumber \\
&&+\frac{J_{H}}{2}\left( \sum_{i,\alpha ,\beta }\Delta _{i}^{\alpha \beta
}C_{i,\sigma }^{\alpha }C_{i,-\sigma }^{\beta }+h.c.-|\Delta _{i}^{\alpha
\beta }|^{2}\right)
\end{eqnarray}%
The physical meaning of all parameters and operators in the
Hamiltonian are presented in order. $H_{0}$ is the noninteracting term \cite%
{arXiv08041113} with
\begin{eqnarray}
\psi _{k,\sigma } &=&(C_{\sigma }^{a}(k),C_{\sigma }^{b}(k)), \\
\epsilon _{k} &=&-(t_{1}+t_{2})(\cos k_{x}+\cos k_{y})-4t_{3}\cos k_{x}\cos
k_{y}-\mu , \\
\gamma _{k} &=&-4t_{4}\sin k_{x}\sin k_{y}, \\
\xi _{k} &=&-(t_{1}-t_{2})(\cos k_{x}-\cos k_{y}),
\end{eqnarray}%
and $\alpha ,\beta =a(xz),b(yz)$ label orbitals, and $\tau $ is the Pauli
matrix for $xz$ and $yz$ orbitals [The superscript $a(b)$ is specified to
``$d_{xz}(d_{yz})$" orbital hereafter]. $H_{I}$ is a Hartree-Fock decomposition
of the on-site interaction term, result from
\begin{eqnarray}
H_{I} &=&U\sum_{i,\beta }n_{i,\uparrow }^{\beta }n_{i,\downarrow }^{\beta }
\nonumber \\
&&+\sum_{i}\left[ U^{\prime }n_{i}^{a}n_{i}^{b}-J_{H}\left( \mathbf{S}%
_{i}^{a}\cdot \mathbf{S}_{i}^{b}+\mathbf{\eta }_{i}^{a}\cdot \mathbf{\eta }%
_{i}^{b}\right) \right] ~,
\end{eqnarray}%
in which $J_{H}$ denotes the Hunds rule coupling, $U(U^{\prime })$ the
on-site Coulomb interaction between electrons on the same (distinct) bands,
and $S_{i}^{\beta +}=C_{i,\uparrow }^{\beta \dag }C_{i,\downarrow }^{\beta
},S_{i}^{\beta z}=(n_{i,\uparrow }^{\beta }-n_{i,\downarrow }^{\beta
})/2,\eta _{i}^{\beta +}=C_{i,\uparrow }^{\beta \dag }C_{i,\downarrow
}^{\beta \dag },\eta _{i}^{\beta z}=(n_{i}^{\beta }-1)/2$. Both $S$ and $%
\eta $ satisfy SU(2) Lie algebra, and $U=U^{\prime }+J_{H}$ \cite{Castellani}%
. Local order parameters $m_{i}^{\beta }$ and $\rho _{i,\sigma }^{\beta }$
are defined as
\begin{eqnarray}
m_{i}^{\beta } &=&\frac{1}{2}\left( \langle n_{i,\uparrow }^{\beta }\rangle
-\langle n_{i,\downarrow }^{\beta }\rangle \right) , \\
\rho _{i,\sigma }^{\beta } &=&\langle n_{i,\sigma }^{\beta }\rangle .
\end{eqnarray}

Respecting to the fact that underline mechanism for superconducting
properties observed in cuprates is controversial, we are not in the position
to discuss what mechanism is for the FeAs-based superconductivity. Instead,
our purpose is to address the interplay between superconductivity and
magnetism so we will start from the assumption that electrons on neighboring
sites pair in the following form

\begin{eqnarray}
H_{SC} &=&-V\sum_{i,\delta ,\beta ,\sigma }\left[ \Delta _{i,\delta ,\sigma
}^{\beta }C_{i+\delta ,\sigma }^{\beta }C_{i,-\sigma }^{\beta
}+h.c.-\left\vert \Delta _{i,\delta ,\sigma }^{\beta }\right\vert ^{2}\right]
\\
&&-V\sum_{i,\delta ^{\prime },\alpha ,\beta ,\sigma }\left[ \Delta
_{i,\delta ^{\prime },\sigma }^{\alpha \beta }C_{i+\delta ^{\prime },\sigma
}^{\beta }C_{i,-\sigma }^{\alpha }+h.c.-\left\vert \Delta _{i,\delta
^{\prime },\sigma }^{\alpha \beta }\right\vert ^{2}\right] ~,  \nonumber
\end{eqnarray}%
where $V$ accounts for the attractive interaction between electrons on
neighboring sites, $\delta (\delta ^{\prime })$ denotes the (next) nearest
neighboring sites [fig. \ref{hopping} (Right)]. 

\begin{figure}[tbp]
\includegraphics[bb=20 320 550 770, width=8.3cm, clip] {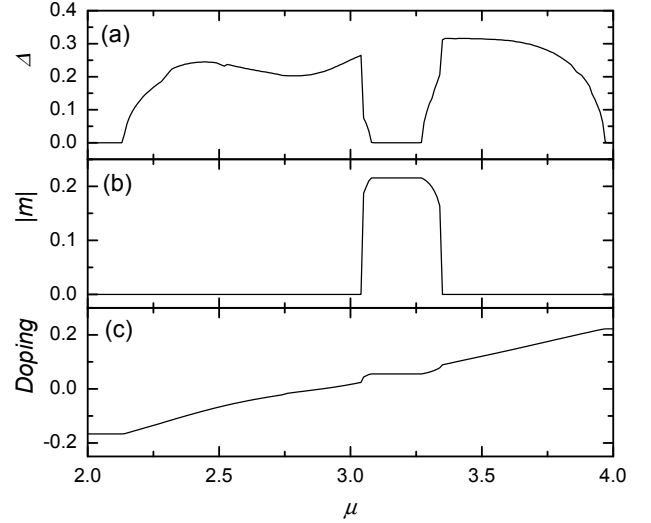}
\caption{ (color online) The superconducting order, the magnetism order, and
doping as function of the chemical potential.}
\label{Miu}
\end{figure}

The mean-field Hamiltonian is quadratic and can be diagonalized numerically
by solving self-consistent equations following standard procedure. In our
numerical calculations, we set $t_{1}=-1$, $t_{2}=1.3$, $t_{3}=t_{4}=-0.85$,
and $J_{H}=1.4\sim 2.0$ in unit of $|t_{1}|$, according the LDA
calculations, and vary other parameters to explore the ground-state
prosperities. Solving such multi-variable nonlinear equations is a very
complicated and time consuming task because of many quantities under
iteration: we have 8 independent $\Delta _{i,\delta ,\sigma }^{\beta }$s, 16
$\Delta _{i,\delta ^{\prime },\sigma }^{\alpha \beta }$s, 4 $\Delta
_{i}^{\alpha \beta }$s, and 4 density parameters $\rho _{i,\sigma }^{\beta }$%
s for each site, and for a $L\times L$-site system, so the total number of
iterated quantities is $32L^{2}$. The reason for us to perform the
unrestricted Hartree-Fock calculations, rather than the restricted ones
(where one would have much reduced number of self-consistent variables), is
to obtain unbiased ground state pairing and magnetic ordering patterns.

To address the competition between the magnetism and superconductivity. We
first define the superconducting and magnetic order parameters in $k$-space
as follows
\begin{eqnarray}
\Delta _{k}^{\alpha \beta } &=&\frac{1}{N}\sum_{jl}e^{i(j-l)k}\Delta
_{j,l}^{\alpha \beta },  \label{eq:superconductingorder} \\
m_{k} &=&\frac{1}{\sqrt{N}}\sum_{j,\beta }e^{ijk}m_{j}^{\beta }.
\label{eq:magneticorder}
\end{eqnarray}
Then we calculate the magnetic order and superconducting order parameter
defined as,
\begin{eqnarray}
\Delta &=&\frac{1}{N}\sum_{\alpha ,\beta ,k}\left\vert \Delta _{k}^{\alpha
\beta }\right\vert , \\
\left\vert m\right\vert &=&\frac{1}{N}\sum_{j,\beta }\left\vert m_{j}^{\beta
}\right\vert .
\end{eqnarray}%
Fig. \ref{Miu} shows the main results of the doping ratio, magnetic order,
and superconducting order as a function of the chemical potential for a
typical parameter set $U=3.4$, $V=0.9$, and $J_{H}=1.5$. Note that paring
driving force usually comes from a second order process so it is expected
that $V$ is smaller than any other renormalized Coulomb interactions. We
firstly observe that there is a flat region in the doping as the chemical
potential varies. Though the position of the flat is a little above the
half-filling for finite size lattices we studied, it will tend to the
half-filling as the system size increases. From fig. \ref{Miu}(b), we notice
that the magnetic order is nonzero around the half-filling. This observation
is consistent with experimental results that the parent compound shows
magnetic order without doping. The dominant magnetic order is $(\pi ,0)$, or
$(0,\pi )$ SDW. Once electrons or holes are doped into the parent compound,
the magnetic order is gradually suppressed to zero, and superconducting
order appears, as shown in fig. \ref{Miu}(a). At half-filling, the
superconductivity is absent. In experiments, the parent compound is a poor
metal without superconductivity. Therefore, our numerical results are
consistent with the experimental results qualitatively. As electrons are
doped into the material, the superconducting order appears at a certain
doing concentration. As the doping concentration becomes larger, it will
disappear again. Similar behavior also occurs in the hole doping region. One
of the most interesting observation is that, in the electrons doping region,
both the magnetism and superconducting orders do not vanish at a small
region. Therefore, the magnetism and superconductivity are not exclusive
with each other. So though there is a competition between two orders, the
magnetism and superconductivity might coexist in a very narrow region.

\begin{figure}[tbp]
\includegraphics[bb=35 353 555 740, width=8.3cm, clip] {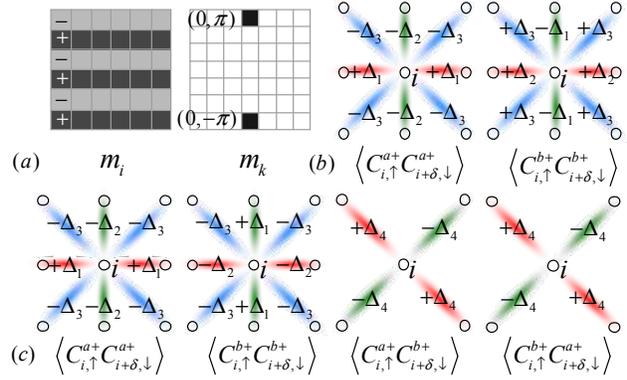}
\caption{ (color online) (a) $(0,\protect\pi)$ spin density wave at
half-filling. (b)Pairing structures of the (next) nearest neighboring sites in
electron doped region. (c) Pairing structures of the (next) nearest neighboring
sites in hole doped region. Here $\Delta_i, i=1,2,3,4$ denote simply the
pairing amplitude which will be shown in main text.} \label{pairing}
\end{figure}

Next we check the magnetic order and pair symmetry of the superconductivity.
The order parameters defined in $k$-space, i.e. eq. (\ref%
{eq:superconductingorder}) and eq. (\ref{eq:magneticorder}), can help us to
learn the details of magnetic order and pairing symmetry. In this work, we
consider three typical cases below.

\emph{Case A (undoped compound):} We consider the magnetic property of the
parent compound for the parameter set
\begin{equation}
U=3.5,V=1.0,J_{H}=1.5,\mu =3.3.
\end{equation}%
Here, we set the initial value of local magnetization at random and let
self-consistent equations converge. We observe that the dominant
magnetization patten in $k$ space include either $(0,\pi )$ or $(\pi ,0)$.
Fig. \ref{pairing}(a) shows the magnetic ordering in both the real and $k$%
-space. In the real space, the compound shows a stripe-like phase which
represents the spin collinear order. The local magnetization are $\pm 0.227$
for light and dark gray respectively. This observation is consistent with
the experimental results of the magnetic properties in undoped case. If we
make a Fourier transformation, we find a spin-density wave of $(0,\pi )$
mode in $k$ space. Therefore, $m_{k}=0.227$ at $(0,\pm \pi )$, and zero
elsewhere. Here, we would like to point out that $(\pi ,0)$ mode can also be
observed even we started iteration with initial conditions consist with of
other order parameters. Moreover, our numerical simulations show that the
magnetic order dominates in the low-doping region.

\emph{Case B (electron doping):} Electron doping typically occurs at
\begin{equation}
U=3.5,V=1.0,J_{H}=1.5,\mu =3.7~,
\end{equation}%
where the solution converges at 13.8\% electron doping. In this large doping
region, the anti-ferromagnetism of the parent compound are suppressed by
superconductivity. The intra-orbital pairing structures of the (next)
nearest neighboring sites are illustrated in fig. \ref{pairing}(b)
schematically. In fig \ref{pairing}(b), $\Delta_{1} = \langle C_{i,\uparrow
}^{a \dagger} C_{i+x,\downarrow}^{a \dagger} \rangle = 0.04824$, $\Delta_{2}
= \langle C_{i,\uparrow }^{a \dagger} C_{i+y,\downarrow}^{a \dagger} \rangle
= 0.05176$ and $\Delta_{3} = \langle C_{i,\uparrow }^{a \dagger}
C_{i+x+y,\downarrow}^{a \dagger} \rangle = 0.03584$. While the inter-orbital
pairings are found to be zero within numerical accuracy. Hence from
numerical results, the nearest-neighbor intra-band pairing belongs to d$%
_{x^{2}-y^{2}}$-wave symmetry ($\cos k_x -\cos k_y$), while the
next-nearest-neighbor intra-band pairing looks like an extended s$_{xy}$%
-wave ($\cos k_x \cos k_y$), and there is no inter-band (next) nearest
neighboring pairing. In addition, the superconducting order parameters in $k$
space obey the following relation
\begin{equation}
\left\langle C_{k,\sigma }^{\beta \dag }C_{-k,-\sigma }^{\alpha \dag
}\right\rangle =\left\langle C_{-k,\sigma }^{\beta \dag }C_{k,-\sigma
}^{\alpha \dag }\right\rangle ,  \label{cccck}
\end{equation}%
which concludes that the pairing symmetry involves spin singlet, orbital
symmetric, and mixed s$_{xy}$ + d$_{x^{2}-y^{2}}$ wave (even parity).

\emph{Case C (hole doping):} Hole doping typically occurs at
\begin{equation}
U=3.5,V=1.0,J_{H}=1.5,\mu =2.8~,
\end{equation}%
where the solution converges at 11.5\% hole doping and without spin density
wave. The first two graphs of fig. \ref{pairing}(c) show that the (next)
nearest intra-band pairing are still mixed s$_{xy}$ + d$_{x^{2}-y^{2}}$
wave. The pairing amplitude are $\Delta_{1} = 0.04525$, $\Delta _{2} =
0.01864$ and $\Delta_{3} = 0.03444$. Moreover, unlike case B, the next
nearest inter-band pairing interactions are not too small to ignore. As
illustrated by the last two graphs in fig. \ref{pairing}(c), the signs of
the order parameters in real space denote the d-wave inter-band pairing
structures. The pairing amplitude is $\Delta_{4} = \langle C_{i,\uparrow
}^{a \dagger} C_{i+x+y,\downarrow}^{b \dagger} \rangle = 0.00185$. We also
notice that the relation $\left\langle C_{k,\sigma }^{\beta \dag
}C_{-k,-\sigma }^{\alpha \dag }\right\rangle =\left\langle C_{-k,\sigma
}^{\beta \dag }C_{k,-\sigma }^{\alpha \dag }\right\rangle $ still holds in
the hole doping region. Therefore, the pairing symmetry are the same as in
case B.

In summary, applying the numerical unrestricted Hartree-Fock approach to a
two-orbital model, we have studied the competition between magnetism and
superconductivity, as well as superconducting pair symmetry in the newly
discovered FeAs-based superconductors. We found that there does exist
competitions between magnetic and superconducting orders. Despite of this,
the magnetism and superconductivity may still coexist in a small region. In
order to find the possible magnetic order and pairing symmetry. We further
studied three different cases, corresponding to undoped, electron doped, and
hole doped, respectively. We found that, around updoped region, the parent
compound shows a spin-density wave of $(0,\pi )$ mode, while in both the
electron and hole doped region, the superconducting pairing symmetry
involves spin singlet, orbital symmetric, and mixed s$_{xy}$ + d$%
_{x^{2}-y^{2}}$ wave(even parity). We noticed also that the result is
somehow different from that of numerical renormalization calculation \cite%
{DHLee08053343}. This might be due to the different range of interactions,
as well as different approach used. Finally, we remark that although we used
the two-orbital model, suggested by previous studies, as the starting model
in our calculation, conclusions drawn should not be altered qualitatively
when more orbitals are included in our approach.

\acknowledgments We are grateful to Dr. Yi Zhou for valuable comments. This
works is supported by the Earmarked Grant for Research from the Research
Grants Council of HKSAR, China (Project No. CUHK 402205).

\end{document}